\documentclass[12pt,preprint]{emulateapj}
\newcommand{\grs}{GRS~1915+105~}
\newcommand{\myemail}{jsyadav@mailhost.tifr.res.in}

\slugcomment{Accepted for publication in ApJ; March, 2006}
\shorttitle{accretion disk and superluminal jets}
\shortauthors{Yadav J. S}

\usepackage{graphicx}
\begin{document}


\title{Connection between accretion disk and superluminal radio jets and
    the role of radio plateau state in \grs}


\author{J. S. Yadav\altaffilmark{1}}
\affil{Tata Institute of Fundamental Research, Homi Bhabha Road,
    Mumbai-400005, India}
\altaffiltext{1}{E-mail: \myemail}

\begin{abstract}
 We investigate the association between the accretion disk 
during radio plateau state  and the following  
superluminal relativistic radio jets with  peak intensity varies from 200 mJy 
to  1000 mJy
observed over a period of five years and present  
the evidences of direct 
accretion disc-jet connection in microquasar \grs. 
We have analysed RXTE PCA/HEXTE X-ray data  and have found that  
the  accretion rate, $\dot{m}_{accr}$, as inferred from the X-ray 
flux, is very high during
the radio plateaux. We   suggest that 
the accretion disk during the radio plateaux  always associated with 
radiation-driven  wind which is manifested in the form  of enhanced 
absorption column density for X-ray and the depleted IR emission. 
We find that the wind density increases with the  accretion disk luminosity
during the radio plateaux.  The wind density is similar to the density of the
warm absorber proposed in extragalactic AGNs
and Quasars. We suggest a simple model for the origin of superluminal 
relativistic jets. Finally, We discuss the implications of this work for 
galactic  microquasars and the extragalactic AGNs and Quasars.
\end{abstract}



\newpage



\section{Introduction}

Large superluminal outflows/jets which appear to move faster than light 
are widely studied in active galactic nuclei (AGN) \& quasars,
yet remain poorly understood (first AGN was discovered in 1943 by
Carl Seyfert!). This is due to the long time scales associated with these 
systems as well as the central part of these systems is  usually obscured by large
amount of dust. Microquasars in our galaxy are stellar mass analogs  of the
massive black hole systems in  quasars and AGNs. Microquasars  are 
closer, smaller and show faster variability that is easily observable.
\citet{mira94} discovered first microquasar \grs in our galaxy  with superluminal 
jets. The black hole mass in \grs is  determined to be 
14$\pm$4 M$_{\odot}$ \citep{grei01}. This microquasar shows exceptionally
high variability in both X-rays and radio. Since 1996, rich X-ray 
variability of this source is observed by RXTE \citep{morg97,muno99} and 
by Indian X-ray Astronomy Experiment, IXAE \citep{yada99}.  \citet{bell00}
have classified the complex X-ray variability of \grs in 12 separate classes on 
the basis of their light curves and the color-color diagrams and suggested
three basic states of this source namely low hard state,  high soft state 
and low soft state.

\begin{figure*}
\includegraphics[angle=270,scale=0.7]{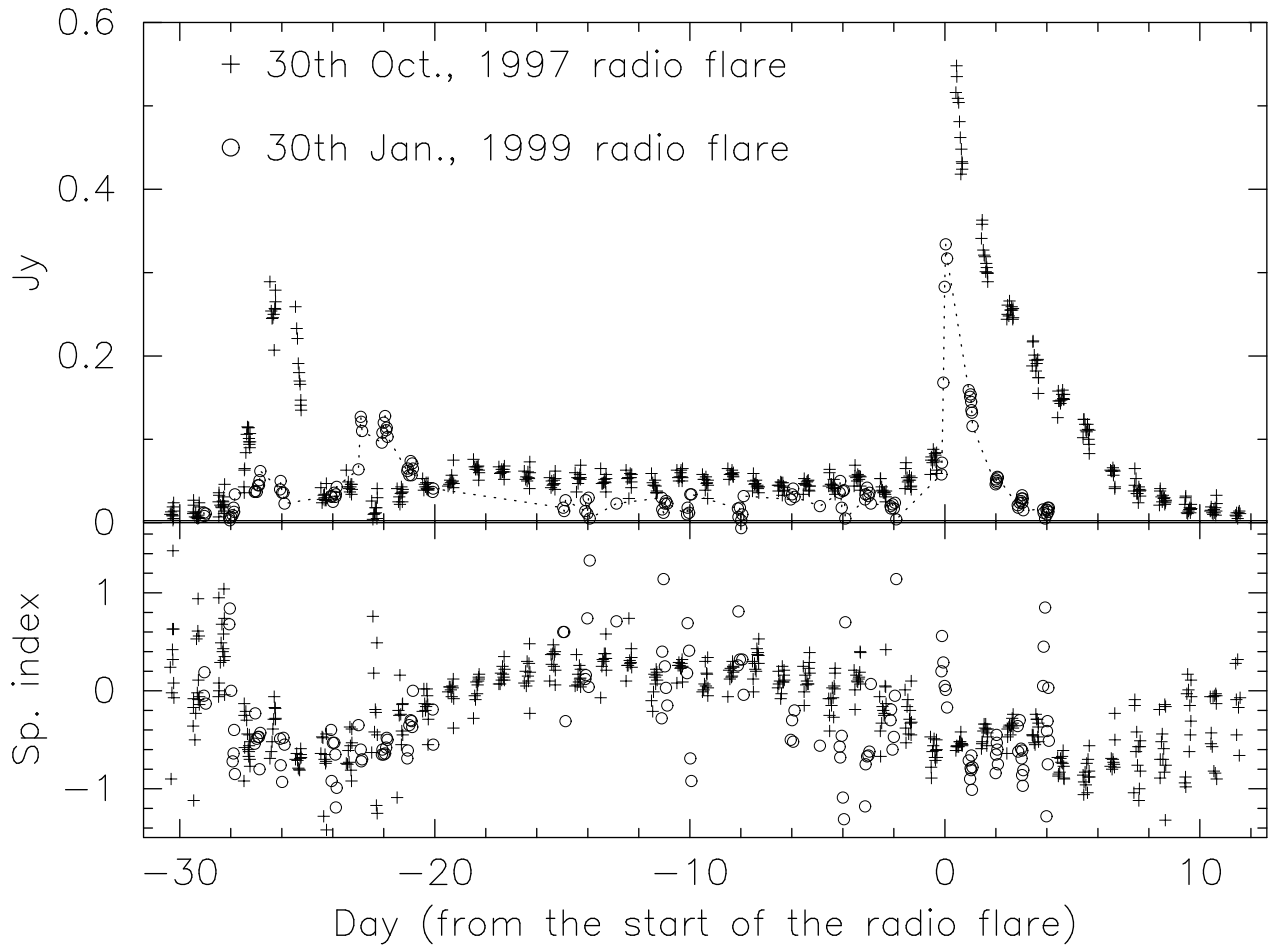}
\caption{GBI 2.25 GHz radio data for 1999 January 30 (start at MJD 51204.7)
and 1997 October 30 (start at MJD 50750.6) large superluminal radio flares
in the top panel.
Start of the flares is offset to 0. It includes preplateau flares,
plateau state and the large superluminal radio flare. The dotted line
connects the data points of 1999 January 30 superluminal radio flare to
guide eyes. The radio spectral index is shown in the bottom panel. }
\label{morph}
\end{figure*}

	The radio emission from \grs can be broadly classified into three
types: (a) steady radio jets (radio plateau state), (b) oscillations/baby
jets (discrete jets) of 20-40 min duration in infrared (IR) and radio, and
(c) large superluminal  radio jets. The steady radio jets of 20 -- 160
mJy flux density are associated with the canonical low hard X-ray state
and observed for extended durations \citep{muno01,fuch03}. These are  
optically thick compact jets of $<$ 200 AU with  velocity $\beta$ of 0.1--0.4 
\citep{dhaw00,ribo04}. The radio emission is correlated  with the X-ray
emission as L$_{radio}$ $\propto$ L$_{X}^{0.7}$ for several different 
sources \citep{gall03}. \citet{pool97} observed  radio oscillations with
delayed emission at  lower frequency. Simultaneous X-ray, IR and radio 
multi-wavelength observations provided first major step in our understanding
of disk-jet interaction and suggested that the spike in X-ray  coincides 
with the beginning of IR flare and it has self absorbed synchrotron 
emission associated with adiabatic expansion \citep{eike98,mira98}.
These are also compact jets with velocity $\beta$ $\sim$ 1 
\citep{dhaw00}. It is further suggested that the spike in X-ray is 
associated  with the change of the X-ray state due to major ejection
episode \citep{yada01}. This is consistent with the suggestion that it
is the coronal material and not the disk material (or any other accretion flow 
associated with the low hard state)
which is ejected prior to radio flares \citep{rau03, vada03,fend04,roth05}.

The relativistic superluminal jets with up to 1 Jy flux density have steep radio
spectrum and are observed at large distances  few hundred AU to 5000 AU from
the core \citep{mira94,fend99,dhaw00}. These radio jets are very energetic with
luminosity close to the Eddington luminosity, L$_{Edd}$ and have been observed
in several sources \citep{fend99,hjel95,wu02,oros01}. Progress in our understanding
of these jets specially their connection to  accretion disk has been slow. A 
semi-quantitative model is proposed for the disk-jet coupling 
\citep{fend04,fend04b}
which although puts all available observational details in perspective but
the scope to gain new insight or new information about disk-jet interaction
seems limited \citep{remi05}.
  The physical connection between X-ray emission  and the superluminal flares
  has been  the hardest to understand. However a general trend is emerging 
  that the plateau states are  generally followed (and perhaps preceded) by
  superluminal flares \citep{fend04b}. 

In this paper, we investigate the association of large superluminal 
jets with the 
radio plateaux. We have analysed the available RXTE PCA/HEXTE X-ray data
during radio plateaux and the radio flare data from the Green Bank 
Interferometer (GBI). We describe our selection criterion and discuss in
detail the morphology of superluminal radio flares. We present tight 
correlations between parameters of the accretion disk and the superluminal jets.
We suggest that the accretion disk during radio plateaux always associated 
with radiation-driven wind  and corroborate it with
supporting evidences. Finally we provide a simple model for the origin of
superluminal jets and discuss its implications for galactic \& extragalactic
radio jet sources.

\section{Observations and analysis}

  All the radio flares observed in \grs  can be
  broadly put into two groups on the basis of their flux, radio 
  spectrum and spatial distribution; (1) the superluminal  flares
  (200--1000 mJy) which have steep radio spectra and are seen at 
  large distances ($\ge$ 240 AU), and (2) all other flares (5--360 mJy) 
  which include the preplateau flares, radio oscillations \& 
  discrete flares  and the steady radio flares 
  during the plateaux. All these flares  have flat radio spectra
  and are observed close to the compact object ($<$ 200AU). It is the 
  superluminal flares which are hard to understand. The other flares
  in the latter group are studied in detail and understood fairly well
  (this is discussed in the last section).
In the internal shock model for superluminal radio jets in microquasars, an
exponential decay is suggested  once the shock is fully 
developed \citep{kais00}.
It is also found that the radio plateau is always associated with superluminal
radio  flare \citep{fend99,klei02,vada03}. We searched 2.25 GHz GBI radio 
monitoring data during the period from 1996 December to 2000 April and 
selected radio plateaux and the following radio flares 
which decay exponentially.
  The preplateau radio flares do not follow exponential decay and are
  like the oscillations and  discrete flares but closely spaced in time as 
  described in the next paragraph.
Seven flares are found to satisfy our  criteria for superluminal flares and are 
listed in Table~\ref{radio_sf}.
One more radio flare on 2001 July 16 which was observed by the Very Large 
Baseline Array (VLBA) and Ratan radio telescope is also 
added in Table~\ref{radio_sf}.
VLBA observations clearly showed an ejecta well separated 
from the core \citep{dhaw03}.
This plus  two more flares out of eight flares listed in Table~\ref{radio_sf} show
moving ejecta  well separated from the core \citep{fend99,dhaw00}. 
In 
Table~\ref{radio_sf}, rise time is the total time taken by a radio flare to rise from
plateau flux to the peak flux. The gaps in the GBI data put upper limit 
on the rise time ($<$ 1 day). For 340 mJy flare on 1999 January 30 the 
rise time 
is 0.25 day while for 710 mJy flare on 1998 June 3 the rise time is 0.3 day. 
These 
values are consistent with the reported values of 0.25 -- 0.5 day for different 
microquasars \citep{fend04}. 
The decay time constant is calculated by fitting
exponential decay profile to the radio flare data. The contribution due to 
continuous radio plateau (like  flares on 1998 April 13 \& 30) or any other
minor flare if any is removed prior to  profile fitting. 
The decay time constant varies from 1.12 to 4.02 days. 
The superluminal radio flare profile can be described as fast rise and 
exponential decay (FRED). Here fast rise (0.2-0.5 day) is in comparison to
the decay time. For comparison the oscillations/discrete radio jets have 
rise and decay time in the range of 0.1 -- 0.3 hr \citep{mira98,ishw02}. 
The integrated radio flux is calculated by integrating the fitted 
exponential function over duration three times the decay time constant.
       
       The typical sequence of events for a superluminal radio flare is 
shown in  Figure~\ref{morph} for 550 mJy radio flare on 1997 October 30
(gap in GBI data at peak) and for 340 mJy radio flare on 1999 January 30.
Start of the radio flares is offset to zero. A superluminal event starts 
with small preplateau flares followed by steady long plateau, followed by
superluminal radio flares. We have described the radio plateau state and 
the superluminal flares  in previous  section. The preplateau flares are studied
in detail using GMRT and RYLE radio telescope data for 2001 July 16 
superluminal flare \citep{ishw02,yada03}. These are  discrete
ejections of adiabatically expanding synchrotron clouds with flat radio 
emission (delayed emission  at lower frequency). They are like the 
oscillations/discrete jets but closely spaced in time and hence produce
overlapped radio flares. Radio emission at 1.4 GHz and 15 GHz supports
flat radio spectrum.

\begin{figure}
\includegraphics[angle=270,scale=0.35]{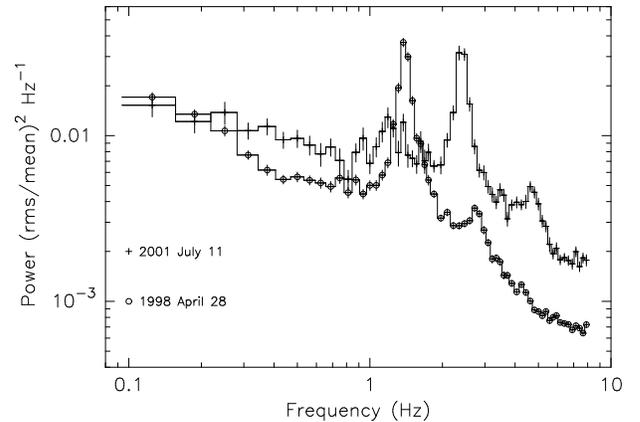}
\caption{Normalised power density spectra in 0.09 -- 9 Hz frequency
range observed on 1998 April 28 and 2001 July 11.  Strong QPOs are seen
at 1.4 Hz and 2.4 Hz in  PDS observed on 1998 April 28 and 2001 July 11
respectively. The first  harmonics are   also visible in spectra.}
\label{qpos}
\end{figure}

	We study X-ray properties during radio plateaux within the 
preceding week from the start of superluminal flares to avoid changes in
accretion disk over the long durations  of plateaux (from -7 to -2  
in Figure~\ref{morph}).  We also avoid last day of plateaux as radio data
suggest rapid changes in  the accretion disk. For the timing analysis, 
we used single-bit-mode 
RXTE/PCA data in the energy ranges 3.6 -- 5.7 and 5.7 -- 14.8 
keV. Power density spectra of 256 bin light curves are generated and
co-added for every 16 s. 
  The power density spectra  are normalised.
  We fit a model of power law + Lorentzian  and have calculated  quasi-
  periodic oscillation (QPO) frequency. 
  We show in Figure~\ref{qpos} 
  the  power density spectra in 0.09 -- 9 Hz frequency      
  range observed on 1998 April 28 and 2001 July 11.  Strong QPOs are seen 
  at 1.4 Hz and 2.4 Hz in  PDS observed on 1998 April 28 and 2001 July 11 
  respectively. The first  harmonics are   also visible in the spectra.
For spectral analysis, we used
PCA standard-2 data (3 --35 keV) and cluster 0 data from HEXTE 
(20 -- 150 keV). We  added 0.5\% systematic error in PCA data. Standard
procedures for data reduction, response matrix and background estimation 
are followed using HEASOFT (ver 5.2). \citet{muno99} have shown that a
spectral fit using sum of multicolor disk blackbody, power law and a 
gaussian line during plateaux results high values of
the inner disk temperature
($\ge$ 3 keV) and unrealistic accretion disk radius ($\le$ 2 km). 
\citet{fuch03} also drew  similar conclusion while analysing X-ray properties
during  a plateau state using INTEGRAL and RXTE data. \citet{rau03} 
analysed large sample of RXTE observations belonging to radio plateaux using
multicolor disk-blackbody + power law spectrum reflected  from the 
accretion disk. This model adequately describes the X-ray spectra of all the
observations, but it yields extremely high values of reflection parameter 
inconsistent with other measurements. 

 A spectral model of simple power law with multicolor disk-blackbody and
 a  gaussian provides reasonable values of temperature and inner radius
 of the disk if the energy spectrum is disk dominated as is the case
 during class $\beta$ \citep{migl03}. During the plateaux, the source is 
 in very high luminosity state (VHS) with power index $\Gamma$ $>$ 2 .
 The spectrum is dominated by the Compton scattered emission ($\ge$ 85\%)
 rather than the disk. The comptonised emission has a complex spectrum 
 shape showing features from both thermal and non-thermal electrons
 \citep{chri05}.   It suggests the presence of an  optically  thick 
 inner disk corona which drains  energy from the disk and reduces the 
 disk temperature. The presence of QPO in 1.4 -- 2.6 Hz frequency range
 is consistent with the large corona (see Table~\ref{radio_sf}).  This 
 will require additional component in the spectral model during the plateaux.
A three component model (multicolor
disk-blackbody + power law + a comptonised component (CompTT)) with a 
gaussian line at 6.4 keV is reported to provide more realistic values of 
the inner disk
temperature and the disk radius \citep{rao00, vada03}. We have used this 
model for our X-ray spectral analysis during plateaux listed in Table~\ref{radio_sf}. 
The absorption column density, N$_H$ is kept as free parameter.  
The integrated X-ray flux for energy range 3 -- 150 keV and 
the derived values of N$_H$ are listed in Table~\ref{radio_sf}. 
The other best-fit X-ray spectral parameters are given in Table~\ref{xray_sf}
along with the flux of the individual spectral components. The 
$\chi^2_{\nu}$ of our spectral fits  falls in the range 0.81 to 1.4 (for
dof  74-94). The spectrum is dominated by the Compton scattering emission
and the contribution from the disk is limited to less than 13\%.
%\citet{vada03}
%have used this spectral model  to study the radio plateaux and
%have given all the fitting parameters which fall within reasonable limits.
%The accretion disc structure during strongly Comptonised spectra (like 
%during VHS) is of great interest and hence we shall discuss these 
%results  separately  \citep{yada05}.

\begin{figure}
\includegraphics[angle=270,scale=0.35]{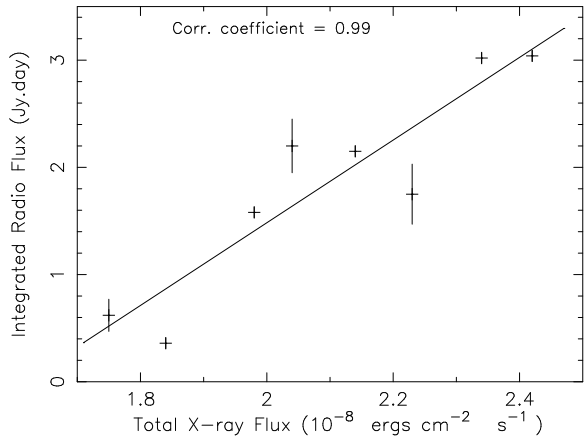}
\caption{Correlation between the total 3-150 keV X-ray flux
from spectral analysis
of RXTE PCA/HEXTE X-ray data during preflare plateau state
and the integrated flux of the superluminal radio flare. The solid line is the
linear fit to the data points.}
\label{xrayradio}
\end{figure}

\begin{figure}
\includegraphics[angle=270,scale=0.35]{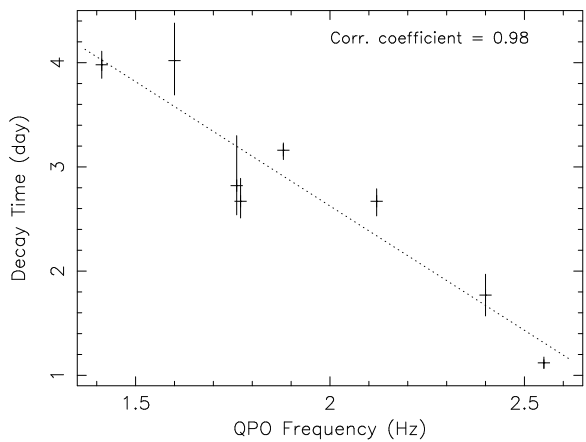}
\caption{Correlation between the quasi-periodic oscillation (QPO) frequency
from timing analysis of  RXTE/PCA
X-ray data during preflare plateau state and the decay time constant of
the superluminal radio flare. The dotted line is the linear fit to the data.}
\label{qpodt}
\end{figure}

\section{Results and discussion}

 It is widely believed that the coronal material (and not the disk 
 material) is ejected prior to radio flares as discussed earlier. It is 
 consistent with the observation that radio flares are observed during 
 transition from the low hard state to the high soft state but never 
 observed during opposite transition \citep{klei02}.  All the  
 radio flares observed  close to the core in \grs can be further 
 subdivided into  two groups on the basis of 
 associated change observed in the accretion disc; (1) the persistent 
 radio flares during the plateaux with steady accretion disk (no change
 associated with radio flares) and, (2) all other  radio flares which
 accompany clear change in the accretion disk. In the former case,  
 inflow and outflow in the accretion disk are in equilibrium. 
 Both the accretion disk and 
 the radio flare are in steady state and  radio emission is tightly 
 correlated  with the X-ray luminosity as discussed above.  In latter case, 
 the hard X-ray component  (Compton component) is either partly or completely
 affected. The large changes will result 
 state transition \citep{roth05,rau03,yada01}. 
 The accretion disk may come back to its initial state after some  time
 or  may remain in the new state for a longer duration. If the accretion
 disk is in some periodic state like in class $\beta$ or $\theta$
 \citep{bell00}, it produces IR and radio oscillations.
 \citet{eike98} 
 suggested that  the spike  in X-ray  during the class $\beta$ is related to 
 mass ejection and the ejected matter decouples from the
  accretion disk. It is further shown that the intensity of these IR and 
  radio jets increases with the  X-ray hardness ratio during the   
  low hard state preceding the spike \citep{yada01}. 
  For these oscillations and the 
  preplateau radio flares, all the available
  experimental data of time  delay in emission at lower 
  frequencies can be adequately
  explained  using adiabatically expanding synchrotron emitting cloud 
  ejected from the accretion disk \citep{ishw02}. These results suggests 
  that the ejected matter behaves like  an adiabatically expanding 
  synchrotron 
  cloud which has been decoupled from the accretion disk.

\citet{muno01}  have shown that radio emission during large superluminal 
flares is decoupled from the accretion disk. This view is further strengthened by
the fact that superluminal flares appear at distances  few hundred AU
from the core \citep{fend99,dhaw00}. 
 \citet{fend99}  have observed    superluminal
 flares after a  radio plateau state.
 To investigate the association of  superluminal flares with plateaux, 
 we plot in Figure~\ref{xrayradio} the total 3--150 keV
 X-ray flux during the preceding plateau vs the integrated radio flux of 
 the superluminal flare. We drive a correlation coefficient of 0.99 suggesting 
a strong connection between the total X-ray flux during the preceding plateau 
state and the integrated radio 
flux of the superluminal flare. We also find strong correlation between ASM
count rate vs  plateau radio flux and ASM count rate vs the integrated radio
flux as discussed by \citet{vada03}(not shown here). In Figure~\ref{qpodt}, 
we plot
QPO frequency from our timing analysis of X-ray data during the plateau state as
a function of decay time constant of the following  superluminal radio flare.
It again shows a tight correlation with correlation coefficient of 0.98. The
remarkable feature of our findings here  is that the parameters calculated using
completely independent spectral and timing analysis  bring out clear 
connection between the accretion disk during the plateau state  and the 
following superluminal radio flare.

\begin{figure}
\includegraphics[angle=270,scale=0.35]{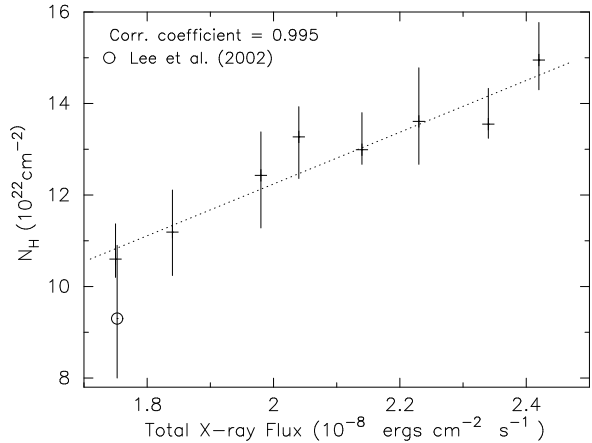}
\caption{Correlation between the total 3--150 keV X-ray flux and the absorption
column density, N$_H$ (plus sign with error bars). The dotted line is the
linear fit to our data. The open circle (with
error bar) is the derived N$_H$ for the solar abundances from CHANDRA and
RXTE X-ray data by Lee et al. (2002) during a plateau state with 20 mJy
radio flux which is similar to the one observed during the plateau preceding
the superluminal flare on 2001 July16 (for detail see text).}
\label{xraynh}
\end{figure}

In the last column of Table~\ref{radio_sf}, we give calculated value of N$_H$ 
which ranges from 10$\times$10$^{22}$ to 15$\times$10$^{22}$ cm$^{-2}$. These 
values are higher than commonly used N$_H$ $\sim$ 5 -- 6$\times$10$^{22}$ 
cm$^{-2}$ for spectral analysis of \grs  by several authors 
\citep{bell97,muno99,yada01,rau03}. Radio emission from atomic hydrogen \& 
molecular hydrogen together puts 
line-of-sight column density $\ge$ 3$\times$10$^{22}$ cm$^{-2}$ 
\citep{dick90,dame01}.
 \citet{bell00} found for the first time evidences for a variable N$_H$
 during the plateaux (class $\chi _1$ \& $\chi _3$) and suggested  
 presence of a large intrinsic absorber. The value of N$_H$ estimated 
 to be  6$\times$10$^{22}$ cm$^{-2}$ for  plateau state while for 
 non-plateau low hard state, it is found to be 2$\times$10$^{22}$ cm$^{-2}$.
% Our spectral model does not overestimate the value of N$_H$. 
% \citet{vada03} have analysed preplateau data of 2001 June 30 when source 
% was in the low hard state (non-plateau, plateau started on 2001 July 03) and 
% estimated  N$_H$ $\sim$ 1.65 $\times$ 10$^{22}$ cm$^{-2}$  which is close 
% to the value estimated by \citet{bell00} for non-plateau low hard state.
\citet{lee02}
have analysed CHANDRA \& RXTE X-ray data during a radio plateau state  with GBI radio  
flux of 20 mJy at 2.25 GHz on 2000 April 24. The total 2--25 keV X-ray flux is 
1.89$\times$10$^{-8}$ ergs cm$^{-2}$ s$^{-1}$. These plateau conditions are identical
(in  radio \& X-ray flux) to the one preceding the superluminal flare on 
2001 July 16 listed in Table~\ref{radio_sf}.
The K absorption edges from Fe, Si and others are detected  in 
CHANDRA data which
suggests the presence of a warm absorber in the vicinity of accretion disk. 
  The N$_H$ derived from Fe absorption edge is found to be 9.3$^{+1.6}_{-1.3}$ 
  $\times$10$^{22}$ cm$^{-2}$ for solar  abundances   which is in agreement
  with our calculated value of N$_H$ $\sim$ 10.6$^{+0.8}_{-0.4}$ 
  $\times$10$^{22}$ 
  cm$^{-2}$ for the superluminal flare on 2001 July 16 (see Figure~\ref{xraynh}). 
The lower limit of bolometric luminosity L$_{bol}$ $\sim$ L$_X$ 
= 6.4$\times$10$^{38}$ ergs s$^{-1}$ which is 0.35 of 
Eddington Luminosity, L$_{Edd}$ for a 
black hole of mass 14 M$_{\odot}$ \citep{lee02}. The corresponding lower limit of bolometric
luminosity for the X-ray flux listed in Table~\ref{radio_sf} falls in the range of 0.35 --
0.48 of L$_{Edd}$.  
\citet{lee02} have suggested   wind from the accretion disk as  the  source of observed
enhanced N$_H$.
One can define wind  (spherical) mass outflow rate, 
$\dot{m}_{wind}$ $\sim$ 4$\pi$r$^2 \rho$v($\Omega/4\pi$) $\sim$ 9.5$\times$10$^{18}$($\Omega/4\pi$)
g s$^{-1}$ where r is the radius (10$^{11}$ cm), $\rho$ is the density of the absorbing
material and v is the wind velocity (100 km s$^{-1}$). The accretion rate, $\dot{m}_{accr}$
= L$_{bol}$/($\eta$c$^2$) $\ge$ L$_X$/($\eta$c$^2$) $\sim$ 7.1$\times$10$^{18}$ g s$^{-1}$ 
where the efficiency $\eta$ $\sim$ 0.1. It shows that as covering fraction ($\Omega/4\pi$) 
approaches unity, the wind outflow rate, $\dot{m}_{wind}$ is comparable to the $\dot{m}_{accr}$.
  During the low hard state, the efficiency of compact persistent radio jets 
  is  estimated to be $\sim$ 5\% \citep{fend01,fend04}. In this case, losses
  due to non-radiative  process (like adiabatic expansion) are likely to 
  dominate. Our estimate of $\dot{m}_{wind}$ above suggests that the  major 
  part of accretion energy goes to the wind during the plateaux. 
When \grs is accreting near  
L$_{Edd}$, the presence of a radiation-driven wind is always expected and the 
wind density should be a strong function of the 
disk luminosity. 
  Our derived values of N$_H$ show strong dependence on observed total X-ray
  flux with correlation coefficient of 0.995 which is shown in 
  Figure~\ref{xraynh}. The dotted line is a linear fit to our data 
  (plus sign with error bars). The open circle with error bar shows the 
  value derived from CHANDRA and RXTE data as discussed above which is in
  agreement with our data within error bars. \citet{kota00}
  have estimated N$_H$ $\sim$ 10$^{24}$ from ASCA data.
%  which is discussed    in next paragraph. 
  Our estimates  of N$_H$ falls between these two values.
%  The N$_H$ increases with  X-ray flux supporting radiation-driven wind
%  as the source of the enhanced N$_H$.

	The best-fit X-ray spectral parameters are given in Table~\ref{xray_sf}
along with the flux of individual spectral components. The spectrum is 
dominated by the Compton scattering emission and the contribution from
the disk is limited  to $<$ 13\%. The inner-disk temperature (kT$_{in}$) and the
plasma temperature (kT$_{e}$) are $\sim$ 0.5 and 4--7 keV 
respectively which are
in acceptable range \citep{vada03}. The derived inner-disk radii  (R$_{in}$) are
in the physically plausible range of (10-25) R$_g$ for 14$\pm$4 M$_{\odot}$
black hole. As the total X-ray flux rises from 1.75$\times$10$^{-8}$ ergs cm$^{-2}$ 
s$^{-1}$ to 2.42$\times$10$^{-8}$ ergs cm$^{-2}$ s$^{-1}$, the major part of it
(close to 80\%) goes to the COMPTT component which shows strong correlation with 
N$_H$ (this is analogous to the solar wind originating from the dense solar corona).
 Our spectral model does not overestimate the value of N$_H$. 
 \citet{vada03} have analysed preplateau data of 2001 June 30 when source 
 was in the low hard state (non-plateau, plateau started on 2001 July 03) and 
 estimated  N$_H$ $\sim$ 1.65 $\times$ 10$^{22}$ cm$^{-2}$  which is close 
 to the value estimated by \citet{bell00} for non-plateau low hard state.
We have also studied the X-ray class $\beta$ in \grs using this spectral model
which results N$_H$ $\sim$ 7$\times$10$^{22}$ cm$^{-2}$ for 
$\chi^2_{\nu}$ close to one \citep{yada06}. Similar N$_H$ has been used earlier
for spectral studies of the X-ray class $\beta$ \citep{migl03}.

Our calculated power density spectra shown in Figure~\ref{qpos} also 
independently lend support to the presence of wind during the plateau state.
In the high soft state, the PDS  is a featureless power law of  
$dP/d\nu \propto \nu^{-\alpha}$ where $\alpha$ is $\sim 4/3$. 
On the other hand, the low hard state is  dominated
by the Compton scattering photon emission and  the PDS develops bandwidth 
limited noise,
a flat-shoulder with a QPO at 1-4 Hz {\citep{rao00b, meie05}. 
As the power of corona
(Compton scattered emission) increases, the power in the PDS increases 
at higher frequency ($\nu \ge$ 0.1 Hz) and the QPO
frequency decreases.  Figure~\ref{qpos} shows PDS spectra observed on 2001 July 11
and 1998 April 28. The Compton scattered flux increases  from 
1.5$\times$10$^{-8}$ ergs cm$^{-2}$ s$^{-1}$ on 2001 July 11 to 
1.9$\times$10$^{-8}$ ergs cm$^{-2}$ s$^{-1}$ on 1998 April 28 while observed
QPO frequency decreases from 2.4 Hz on 2001 July 11 to 1.4 Hz on 1998 April 28.
It is clear from Figure~\ref{qpos} that power at $\nu > 0.2$ Hz is 
less in the PDS observed on   1998 April 28 than that observed  on  
2001 July 11. The fast
variability is suppressed by photon scattering in the enhanced  wind on
1998 April 28 and hence reducing the power in the PDS at higher frequencies.
\citet{shap06} have discussed the decrease in the PDS power observed in 
Cyg X-1 at higher frequency ($\nu >$ 0.1 Hz) as wind increases. In Cyg X-3, 
the PDS power  in the low hard state  drops  to below 10$^{-3}$  
(rms/mean)$^2$ Hz$^{-1}$ at frequencies $\nu  > 0.1$ Hz as dense wind
 from the companion always envelops the compact object \citep{yada06}.

\begin{figure}
\includegraphics[angle=270,scale=0.35]{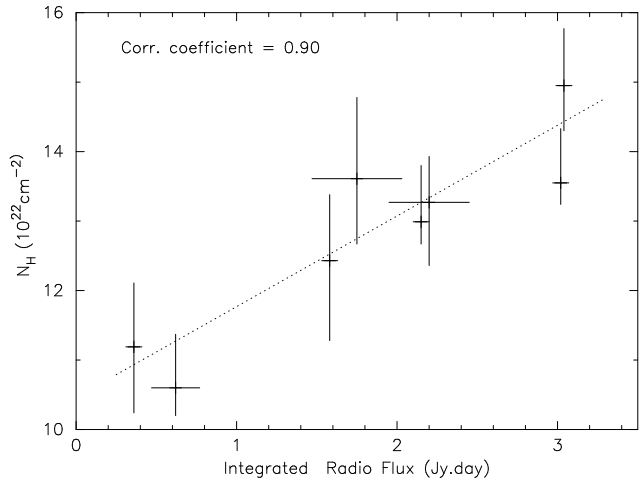}
\caption{Correlation between  the integrated radio flux of the superluminal radio
flares and the  absorption column density, N$_H$ calculated from X-ray spectral
analysis of RXTE PCA/HEXTE data during preflare plateau state.  The dotted line is a
linear fit to the data.}
\label{windradio}
\end{figure}

\citet{kota00} analysed ASCA data  of two \grs observations in the low hard state in 1994 \&
1995  and detected characteristic  absorption  lines which suggest large values of N$_H$.  
They have discussed  radiation-driven wind from the accretion disk as the source of 
enhanced N$_H$
and  estimated  wind terminal velocity of the order of 
10$^8$ cm s$^{-1}$
for the values of r and L$_X$ discussed above.  They have speculated wind's role in 
the formation of 
superluminal radio jets. The spectral analysis of energy spectra during the extended low soft state
requires large N$_H$ which lends further support for the 
radiation-driven wind from the 
accretion disk \citep{vada02}. The evidences for a Seyfert-like warm absorber have been found in 
other microquasars like GRO J1655-40, GX 339-4 and XTE J1650-500 beside \grs \citep{mill04,ueda98,
lee02,kota00} and superluminal flares have been observed  in all these sources except XTE J1650-500.
In Figure~\ref{windradio}, we plot the integrated radio flux of superluminal flare  as a function of N$_H$
calculated from the X-ray data during the preceding plateau state. 
It  brings out  clearly the  connection between the superluminal flare and the enhanced N$_H$ 
with correlation coefficient of 0.90. The dotted line is a linear fit to the data.
The wind density for typical one AU sphere comes out to be  $\sim$ 10$^9$ cm$^{-3}$ which is similar 
to the value proposed in quasars \citep{elvi00}. 

The presence of a dense absorber close to the vicinity of the accretion disk
should affect the IR emission during the radio  plateaux. The source size of 
the steady compact jet during
the radio plateau state varies as a function of frequency which is consistent with self-absorbed 
steady outflow \citep{dhaw00,fuch03,fend04}.  It is similar to the self-absorbed discrete
outflows in the case of oscillations/baby jets which produce delayed emission at lower 
frequencies \citep{mira98,eike98,ishw02,ueda02}. In both the cases, the radio emission is flat or
inverted. The IR emission during radio plateaux in \grs \citep{fuch03a,fuch03} is  
closer to the IR emission in Cyg X-3 where wind from the companion always envelops the 
compact object \citep{ogle01,koch02}.

Our results  support the internal shock model for the origin of superluminal flares 
\citep{kais00}.  The internal shock should form in the previously generated slowly moving
wind from the accretion disk with $\beta$ $\le$ 0.01 \citep{kota00} as the fast moving discrete
jet with $\beta$ $\sim$ 1 \citep{dhaw00} catches up and interacts with it. Both the components;
 slow moving wind and  fast moving jet are related to the accretion disk during  plateau state and 
the strength \& speed of these two components will determine the power of the internal shock. 
This aspect has been brought out clearly by the correlations presented in 
Figures~\ref{xrayradio},~\ref{qpodt} \space \&~\ref{windradio}. In our
case Lorentz factors $\Gamma_2$--$\Gamma_1$ $\sim$ $\Gamma_2$ for these two components \citep{
fend04, rees94}. The wind deposits large amount of energy as $\dot{m}_{wind}$ approaches 
$\dot{m}_{accr}$
prior to the switching on of  superluminal flare. \grs produces upto 1 Jy superluminal flares with
apparent velocity, $\beta_{app}$ in a range 1.2--1.7. It is a efficient  internal shock scenario where 
not just velocity but total jet power has significantly increased \citep{fend04}. This view is
further strengthened by the fact that Cyg X-3 can produce superluminal flares with flux density 
upto 15 Jy with $\beta_{app}$  $\sim$ 0.69  where the companion star can provide denser wind. 
The internal shock model can easily accommodate high  jet power requirement $\ge$ 10$^{38}$ ergs 
s$^{-1}$ \citep{fend99} and can explain the shifting from thick to thin radio emission during
superluminal flares \citep{kais00}.

 Radio plateaux  occur when $\dot{m}_{accr}$, as inferred from the X-ray flux, 
 is very high and L$_{bol}$ approaches  
 L$_{Edd}$. As $\dot{m}_{accr}$ rises, the 
accretion disk passes through an instability zone (prior to plateau state) and produces preplateau
flares (see Figure~\ref{morph}). Thereafter, it enters in steady plateau state. At the end of the plateau
state, changes in $\dot{m}_{accr}$ again produce instabilities in the accretion disk which produce
post-plateau flare (flare which produce superluminal flare) and the source comes out of plateau 
state. It may be noted here that a post-plateau flare may not always end plateau state but it may
change the level of the plateau as is the case of superluminal flares on 1998 April 13 \& 30
(plateau radio flux 91 mJy (highest flux in Table~\ref{radio_sf})). The instabilities in the   
accretion disk may be triggered by  either a change in $\dot{m}_{accr}$ or by some adjustment among
different accretion flows \citep{yada99}. This view is supported by the finding that certain X-ray
variability classes are observed preferably before and after the  plateaux (low hard state)
\citep{naik02}.

Our results presented here suggest only three types of radio emission in \grs; (1) radio 
emission during the steady plateau state with $\beta$ in a range 0.1--0.4 (steady X-ray 
properties), (2)
oscillations or discrete baby jets with $\beta$ $\sim$ 1 (state transition in X-rays), and
(3) faint flares of 2--3 min durations with likely low $\beta$ (hard X-ray spectrum changes 
but no state transition \citep{roth05}). The superluminal flares are the consequences of 
class 1 \& class 2 radio emissions when occur in this order (but not in the reverse order). 
Our description of superluminal flares here can  explain why axis of superluminal flares
differs by $\le$ 12$^o$ from the axis of  AU scale compact jets \citep{dhaw00}. The wind 
is radiation-driven and is expected  to originate from the inner part of the accretion disk while
the compact jets are most likely to originate from the outer corona. 
The phase lag  between the hard and soft
X-ray photons in \grs changes  sign during the plateau state and produce complex behavior
\citep{muno01}. \citet{varn05} has suggested that anything located inside the inner radius of 
the accretion disk (in our case wind) that can absorb a small part of soft X-ray flux can
explain  the changing of sign of the lag and its complex behavior during the plateau state.
Other consequences of our results is that one would not expect oscillations/ baby jet type of
IR \& radio emission in Cyg X-3 (not observed so far \citep{dhaw00}) as has been observed
in \grs \citep{mira98, eike98,ishw02}.
  Any discrete jet (including oscillations) should produce internal shock as 
  wind from the companion is always present in the case of Cyg X-3 and 
  therefore, such jet would always end up producing a superluminal jet 
  with thin radio emission.

The physics  in all the systems dominated  by black holes is essentially 
the same and therefore, it is always tempting to compare microquasars
with massive extragalactic AGNs and quasars.
The correlation between radio and X-ray luminosity during the low hard state  
\citep{gall03} can be extended to the AGNs by including a black hole mass 
term \citep{merl03,falc04}. Therefore  the steady jets in microquasars during
radio plateaux are directly comparable with those in AGNs. \citet{mars02} have
claimed to observe oscillation radio jet ($\sim$ 1/year) in AGN 3C120. 
They related these radio jets to superluminal ejection events  
which can be understood
in the frame work of the model discussed here (if we assume 
the presence of warm absorber).
The results discussed in this paper suggest the propagation of shock wave
to produce the superluminal jets in microquasars, 
a interpretation  usually favoured for
jets in  extragalactic systems. It is shown recently that the jets in 
microquasars  are as relativistic as AGN jets provided  the jets in 
microquasars are not confined and the derived opening angles are solely 
due to the relativistic effect \citep{mill06,gopa04}. It is natural
to associate (though disputed!) the radio loud and radio quiet dichotomy
observed in AGN with jet-producing low hard state and non-jet-producing 
high soft state  in microquasars \citep{macc03}. The PDS of luminous 
narrow line Seyfert 1 AGNs appears to be similar to the one observed 
during the high soft state in microquasars while the PDS of the low 
luminosity AGNs is similar to the one observed during the low hard
state in microquasars \citep{mcha04,mark05}.
Shocks are related to the most
energetic  processes in the universe (Gamma ray bursts,  large superluminal jets
in microquasars and AGNs) and microquasars provide a great opportunity to study 
them under the best possible conditions within reasonable time.

\section{Conclusions}
We present for the first time clear evidences of direct connection between 
the accretion disk during plateaux and the following superluminal radio flares.
We find that the  $\dot{m}_{accr}$, as inferred from the X-ray flux, 
is very high during the radio plateaux and
L$_{bol}$ approaches  L$_{Edd}$.
We suggest that such hot accretion disk during the radio plateaux  always accompany with
radiation-driven wind.  The internal shock  forms in the previously generated slowly moving
wind from the accretion disk with $\beta$ $\le$ 0.01  as the fast moving 
discrete
jet with $\beta$ $\sim$ 1  catches up and interacts with it. Both the components;
slow moving wind and fast moving jet are related to the accretion disk during plateau 
state and
the strength \& speed of these two  should determine the power of the internal shock.
In \grs, it  is a efficient  internal shock scenario where
not just velocity but total jet power has significantly increased.
  The superluminal flares have steep radio spectrum and are observed at 
  large distances from the compact object ($\ge$ 240 AU). The profile 
  of the superluminal radio flares can be described as  fast rise and
  exponential decay (FRED).

  The other radio flares observed in \grs  include the persistent radio 
  emission during the plateaux (20--160 mJy), the  radio oscillations
  \& discrete radio flares (5--150 mJy) and the preplateau radio flares
  (50--360 mJy). All of these are observed  close  to the compact
  object ($<$ 200 AU) and have flat or inverted radio spectra.
  During the radio plateaux, both the radio flare and the accretion
  disk are in steady state (outflow and inflow in the accretion disk are in equilibrium) and 
  a tight correlation is observed between the radio luminosity and the 
  X-ray luminosity as discussed earlier. The IR \& radio oscillations (5--150)
  are periodic ejections of adiabatically expanding  self absorbing 
  synchrotron  clouds \citep{mira98,ishw02}. The IR flux suggests that
  the ejected matter decouples from
  the accretion disk at the time of the spike in the X-rays \citep{eike98}.
  The preplateau radio flares (50--360 mJy) are discrete ejections
  which are closely spaced in time and hence produce overlapped 
  radio flares. The preplateau
  flares have been modeled as  adiabatically expanding 
  self absorbing clouds ejected from the accretion disk which 
  explain  reasonably well all the available observed data of
  time delay for radio emission at lower frequency \citep{ishw02}. 
  It is widely believed that it is the coronal material and not the 
  disk material which is ejected prior to radio flares \citep{rau03, 
  vada03,fend04,roth05}. During oscillations/discrete flares and
  the preplateau flares, the coronal mass ejections  affect the
  hard X-ray component (Compton component) either partly or completely
  (state change) (outflow and inflow are not in equilibrium in this case
  unlike  during the plateaux).
\acknowledgments
Author thanks the referee for his/her useful comments which have greatly
improved  the presentation of the paper and will enhance its impact.
Author  also thanks  RXTE PCA/HEXTE and NSF-NRAO-NASA Green Bank Interferometer 
teams for making 
their data publicly available.  The Green Bank Interferometer is a facility 
of the National Science Foundation operated by the NRAO in support of NASA 
High Energy Astrophysics programs.

\clearpage

\begin{table}[t]
\caption{All selected superluminal radio  flares and their properties (see text for
details). X-ray Properties from RXTE PCA/HEXTE data during preceding
radio plateau.}
\begin{center}
\scriptsize
\begin{tabular}{|ccccc|cccccc|}
\hline
\hline
\multicolumn{5}{|c|}{Superluminal Radio Flare Properties }  &\multicolumn{6}{|c|}{Associated Preceding Plateau Properties} \\
\hline 
MJD/ & Peak&Rise&Decay Time &Radio & GBI &
ASM &Date of &QPO &Total & N$_H$ \\
Date &Flux & Time  &constant &Flux$^a$ &Flux & Flux& RXTE& Freq.&X-ray&(10$^{22}$ cm$^{-2}$) \\
 &(mJy) &(day) & (days) &(Jy.day) & (mJy) &(c/s) &Obser.
&(Hz)&Flux$^b$& \\
\hline
50750/1997 Oct 30  &550 &$<$0.6 &3.16$^{+0.07}_{-0.09}$ &2.20$^\#$ &47.3  &35.9 &1997 Oct. 25&1.88&2.04&13.27$^{+0.66}_{-0.91}$ \\
50916/1998 Apr 13  &920 &$<$0.8 &4.02$^{+0.36}_{-0.33}$ &3.04 &91.0  &48.2 &1998 Apr. 11&1.60&2.42&14.95$^{+0.82}_{-0.65}$ \\
50933/1998 Apr 30  &580 &$<$0.9 &3.98$^{+0.13}_{-0.13}$ &3.02 &91.0  &44.9 &1998 Apr. 28&1.41&2.34&13.55$^{+0.78}_{-0.31}$ \\
50967/1998 Jun 03  &710 &0.3    &2.82$^{+0.48}_{-0.28}$ &2.15 &56.2  &36.5 &1998 May 31 &1.76&2.14&12.99$^{+0.81}_{-0.32}$ \\
51204/1999 Jan 30  &340 &0.25   &1.12$^{+0.03}_{-0.05}$ &0.36 &27.8  &33.8 &1999 Jan. 24&2.55&1.84&11.19$^{+0.92}_{-0.95}$  \\
51337/1999 Jun 08  &490 &$<$0.7 &2.67$^{+0.22}_{-0.16}$ &1.58 &45.5  &35.2 &1999 Jun. 03&1.77&1.98&12.43$^{+0.95}_{-1.15}$  \\
51535/1999 Dec 23  &510 &$<$0.8 &2.67$^{+0.12}_{-0.14}$ &1.75$^\#$ &51.3  &38.9 &1999 Dec. 21&2.12&2.23&13.61$^{+1.17}_{-0.94}$  \\
52105/2001 Jul 16$^c$  &210 &$<$0.7&1.77$^{+0.20}_{-0.20}$ &0.62$^\#$ &20.0  &30.0 &2001 Jul. 11&2.40&1.75&10.60$^{+0.77}_{-0.40}$  \\
\hline
\multicolumn{11}{l}{{\bf a:} Integrated flux is obtained by fitting an exponential function to the flare profile and integrating the function over duration }\\
\multicolumn{11}{l}{~~~~of 3$\times$decay time constant. Typical errors are $\pm$0.05 (larger error (0.15--3.0) when marked ``\#'' due to large gap in GBI radio}\\ 
\multicolumn{11}{l}{~~~~data or due to contribution by additional radio flare)} \\
\multicolumn{11}{l}{{\bf b:} Integrated 3--150 keV X-ray flux in 10$^{-8}$ ergs cm$^{-2}$ s$^{-1}$} \\
\multicolumn{11}{l}{{\bf c:} VLBA radio data (see text for details)} \\
\end{tabular}
\label{radio_sf}
\end{center}
\end{table}

\begin{table}[t]
\caption{Best fit spectral parameters for the X-ray observations prior to the   radio    
flares shown in Table 1.}
\begin{center}
\scriptsize
\begin{tabular}{|ccccccccccc|}
\hline
\hline
Day of &\multicolumn{4}{c}{3 -- 150 keV flux (in 10$^{-8}$ ergs cm$^{-2}$ s$^{-1}$)}&kT$_{in}$
&kT$_{e}$& $\tau$&   $\Gamma$ &R$_{in}^a$ &$\chi^2_{\nu}$(dof)\\
RXTE obs.& F$_{total}$  & F$_{dbb}$ & F$_{ctt}$ & F$_{pow}$       &keV   &keV  &        &          &km       &\\
\hline 
\hline
1997 Oct. 25&2.04&0.20 & 0.75&0.97&0.50$^{+0.02}_{-0.01}$ & 5.50$^{+0.16}_{-0.23}$ & 7.7$^{+0.8}_{-0.6}$ & 2.47$^{+0.30}_{-0.03}$ & 748.1 & 1.40(74) \\
1998 Apr. 11&2.42&0.29 & 0.93&1.05 & 0.46$^{+0.01}_{-0.01}$ & 4.04$^{+0.13}_{-0.30}$ & 9.7$^{+0.9}_{-0.8}$ & 2.97$^{+0.09}_{-0.03}$ & 1392.4 & 0.83(94) \\
1998 Apr. 28&2.34&0.29 & 1.04&0.85& 0.50$^{+0.01}_{-0.01}$ & 4.18$^{+0.15}_{-0.22}$ & 8.8$^{+1.0}_{-0.4}$ & 2.50$^{+0.12}_{-0.19}$ & 831.4 & 1.43(94) \\
1998 May 31 &2.14&0.25 & 0.85&0.91& 0.51$^{+0.01}_{-0.01}$ & 4.96$^{+0.22}_{-0.51}$ & 8.5$^{+1.0}_{-0.7}$ & 2.69$^{+0.24}_{-0.06}$ & 741.7 & 0.93(94) \\
1999 Jan. 24&1.84&0.128& 0.60 &1.0 & 0.57$^{+0.01}_{-0.01}$ & 7.00$^{+0.22}_{-0.60}$ & 4.2$^{+1.9}_{-0.3}$ & 2.39$^{+0.14}_{-0.09}$ & 297.0 & 1.15(93)  \\
1999 Jun. 03&1.98&0.23 & 0.73&0.92& 0.51$^{+0.01}_{-0.01}$ & 5.98$^{+0.80}_{-0.60}$ & 7.7$^{+0.7}_{-1.3}$ & 2.52$^{+0.23}_{-0.12}$ & 660.9 & 1.16(88)  \\
1999 Dec. 21&2.23& 0.25&  0.71& 1.16& 0.48$^{+0.02}_{-0.01}$ & 5.25$^{+0.76}_{-0.39}$ & 8.3$^{+1.1}_{-2.2}$ & 2.82$^{+0.03}_{-0.07}$ & 981.3 & 0.81(88) \\
2001 Jul. 11&1.75& 0.13 & 0.50& 0.95& 0.54$^{+0.02}_{-0.01}$ & 6.99$^{+0.36}_{-0.59}$ & 7.1$^{+0.5}_{-1.5}$ & 2.39$^{+0.03}_{-0.07}$ & 354.3 & 0.98(87) \\
\hline
\multicolumn{11}{l}{{\bf $^a$}: $R_{in}$ derived assuming distance $D=12.5$ kpc and inclination angle, $\theta=70^{o}$}\\
\multicolumn{11}{l}{: associated best fit values of N$_H$ is shown in
     Table 1}\\
\end{tabular}
\label{xray_sf}
\end{center}
\end{table}
\end{document}